\documentclass[proof]{WileyASNA-v1}

\articletype{Article Type}%

\received{XX XXX XXXX}
\revised{XX XXX XXXX}
\accepted{XX XXX XXXX}

\def\feii{\rm{Fe \sc{ii}}}

\def\hb{H$\beta$}
\def\rfe{R$_{\rm{FeII}}$}
\def\RL{$R\mathrm{_{H\beta}}-L_{5100}$}

\def\kms{\,km\,s$^{-1}$}

\def\hb{{\sc{H}}$\beta$\/}

\def\feii{{Fe \sc{ii}}}

\def\rfe{R$_{\rm{FeII}}$}

\def\RL{$R\mathrm{_{H\beta}}-L_{5100}$}

\def\zsun{Z$_{\odot}$}

\def\rblr{$R\mathrm{_{BLR}}$}

\def\hb{H$\beta$}

\def\hb{{\sc{H}}$\beta$\/}

\def\lcont{$L_\mathrm{cont}$}
\def\ledd{$L\mathrm{_{Edd}}$}

\newenvironment{myfont}{\fontfamily{lmtt}\selectfont}{\par}
\DeclareTextFontCommand{\textmyfont}{\myfont}

\begin{document}

\title{Taming the derivative: diagnostics of the continuum and \hb{} emission in a prototypical Population B active galaxy}

\author[1,2,3]{Swayamtrupta Panda*}

\author[4]{Edi Bon}

\author[5]{Paola Marziani}

\author[4]{Nata{\v s}a Bon}

\authormark{Panda \textsc{et al}}

\address[1]{\orgname{Center for Theoretical Physics, Polish Academy of Sciences}, \orgaddress{Al. Lotnik\' ow 32/46, 02-668 Warsaw, \country{Poland}}}

\address[2]{\orgname{Nicolaus Copernicus Astronomical Center, Polish Academy of Sciences}, \orgaddress{Bartycka 18, 00-716 Warsaw, \country{Poland}}}

\address[3]{\orgname{Laborat\'orio Nacional de Astrof\'isica - MCTIC}, \orgaddress{R. dos Estados Unidos, 154 - Na\c{c}\~oes, Itajub\'a - MG, 37504-364, Brazil}}

\address[4]{\orgname{Astronomical Observatory Belgrade}, \orgaddress{Volgina 7, 11060 Belgrade, \country{Serbia}}}

\address[5]{\orgname{National Institute for Astrophysics (INAF), Astronomical Observatory of Padova}, \orgaddress{Vicolo dell'Osservatorio, 5, 35122 Padova PD, \country{Italy}}}

\corres{\email{panda@cft.edu.pl}}

\abstract{We report preliminary results on the analysis of the continuum and \hb{} light curves of the type-1 active galactic nucleus (AGN) NGC 5548. We notice a clear signature of shallowing in the trend between the \hb{} and the continuum luminosities. We attempt the recovery of this observed \hb{} emission trend as a response to large continuum flux increase using \textmyfont{CLOUDY} photoionization simulations. We explore a wide range in the physical parameters space for modelling the \hb{} emission from the broad-line region (BLR) appropriate for this source. We employ a constant density, single cloud model approach in this study and successfully recover the observed shallowing of the \hb{} emission with respect to rising AGN continuum. With our modelling, we are able to provide constraints on the local BLR cloud density and recover the BLR distances (from the continuum source) consistent with the \hb{} reverberation mapping estimates. We further discuss the implications of the BLR covering factor and their sizes on recovering the observed trend.}


\jnlcitation{\cname{%
\author{Panda S.}, 
\author{Bon E.}, 
\author{Marziani P.}, and 
\author{Bon N.}} (\cyear{2021}), 
\ctitle{Taming the derivative: diagnostics of the continuum and \hb{} emission in a prototypical Population B active galaxy}, \cjournal{Astronomische Nachrichten}, \cvol{2021;00:1--6}.}

\maketitle

\footnotetext{\textbf{Abbreviations:} AGN, Active Galactic Nuclei; NGC, New General Catalogue; BH, Black Hole; Mrk, Markarian; MJD, Modified Julian Day; STORM, Space Telescope and Optical Reverberation Mapping; SAO, Special Astrophysical Observatory; ULySS, University of Lyon Spectroscopic analysis Software; 2MASS, 2 Micron All Sky Redshift Survey; SED, Spectral Energy Distribution; LOC, Locally Optimized Cloud; CF, Covering Factor}

\section{Introduction}\label{sec1}

NGC 5548 has been hailed as an archetypical type-1 AGN \citep{ursinietal15}, and serves as a valuable laboratory to study the long-term variation of the broad-line region \citep[BLR][]{wanders1996,sergeev2007} and reliability of the reverberation-mapped black hole (BH) mass measurements \citep{Petersonetal2002, collin2006}. It has been the target of many reverberation campaigns, notable among them were the International AGN Watch and AGN STORM programs \citep[see][]{Petersonetal2002,derosaetal2015}. These campaigns successively have provided information on the geometry, ionization structure, and kinematics of the broad-line emitting regions in this source \citep[][and references therein]{derosaetal2015,horneetal2021}.

Another way to look at the evolution of this source is tracing it along the optical plane of the Eigenvector 1 schema. The Eigenvector 1 schema dates back to the seminal work by \citet{bg92}, wherein the authors performed a principal component analysis using observed spectral properties for a sample of 87 low-redshift sources to realise the, now well-known, main sequence of quasars. The Eigenvector 1 schema, or the optical plane of the main sequence of quasars, i.e. the plane between the FWHM of broad \hb{} emission line and the strength of the optical \feii{} emission\footnote{the \feii{} emission is the integrated emission within 4434-4684 \AA.} with respect to \hb{}. This has been a key subject of study spanning close to 3 decades that has advanced our knowledge of the diversity of Type-1 AGNs both from observational and theoretical aspects \citep[][and references therein]{bg92,sulenticetal00a,sh14,sul15,mar18,panda18b,panda19b}.  In the Eigenvector 1 main sequence, NGC 5548 appears as a prototypical Population B source \citep{sulenticetal00a}, where Population B includes sources radiating at modest Eddington  ratio ($\lesssim 0.2$).  NGC 5548 shows significant variability in the optical and UV bands of the electromagnetic spectrum, and has been the target of over a dozen reverberation mapping campaigns \citep[see][and references therein]{Luetal2016}. In addition to a prototypical spectrum, the source makes transitions from low to high states yet stays within the Population B class in terms of FWHM \hb\ and \rfe{} \citep[we refer the readers to Figure 2 of][where they show the overall movement of the source on the Eigenvector 1 plane]{bonetal2018}. In this context, NGC 5548 presents several intriguing variability patterns that is inclusive of the changes observed in the Balmer lines, primarily H$\alpha$ and \hb{}.

\begin{figure}
    \centering
    \includegraphics[width=\columnwidth]{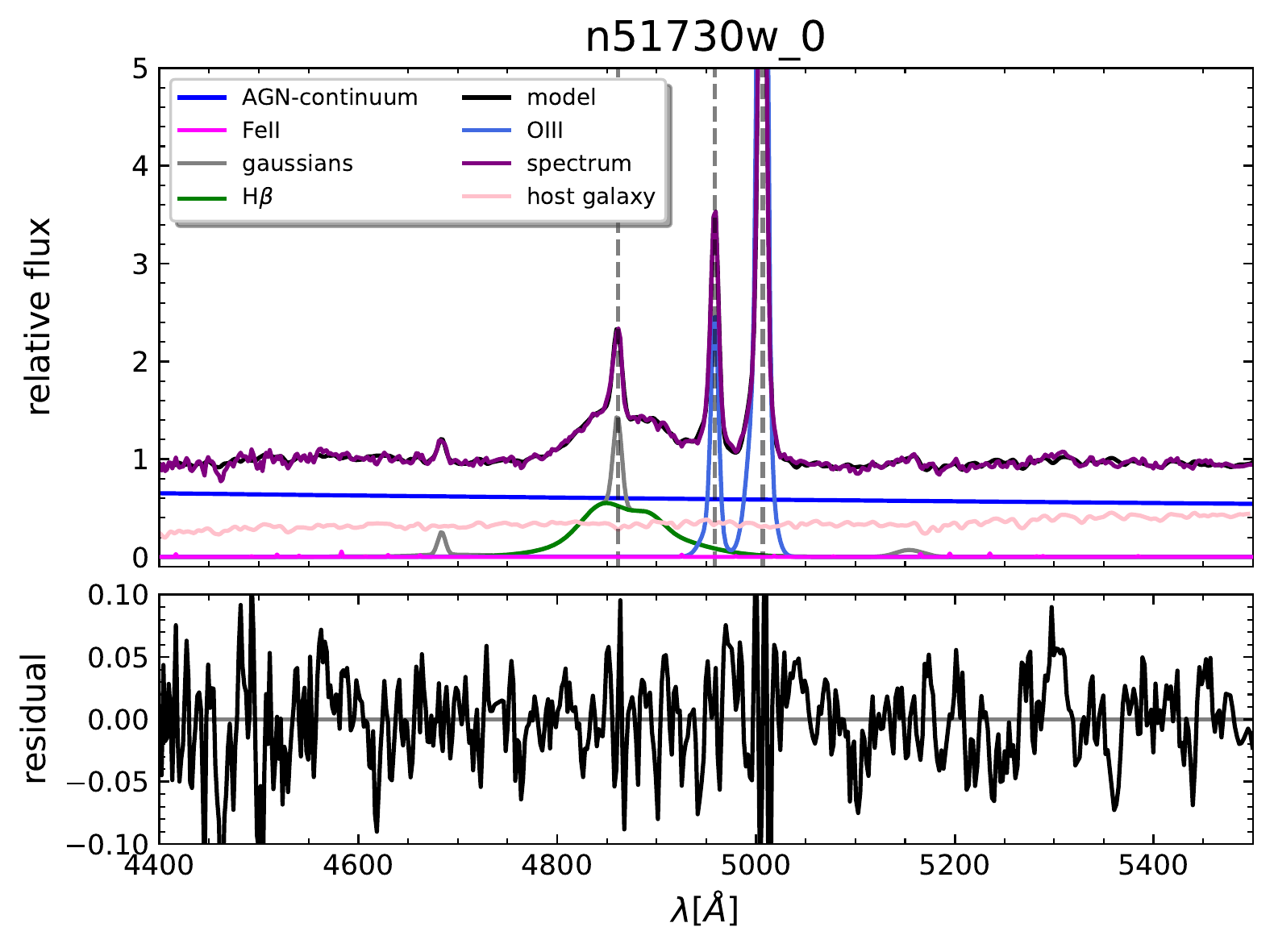}
    \caption{Spectral decomposition shown for a representative spectrum for NGC 5548 used in this work. The spectrum was observed on 51730 MJD. Lower panel shows the residual for the observed spectrum (shown in purple) and the modelled spectrum (shown in black) over the wavelength range. The various components depicted in the figure are: AGN power-law continuum (in dark blue), \feii{} pseudo-continuum (in magenta), extra gaussians (in gray), \hb{} (in green), overall model (in black), [O III] doublet (in blue), observed spectrum (in violet), and modelled host galaxy contribution (in pink).}
    \label{fig:spectra}
\end{figure}

In this paper, we test the connection between the continuum variability in the optical regime and the corresponding \hb{} response to it for NGC 5548, in order to realize the increase, albeit with a gradual saturation, in the \hb{} emitting luminosity with increasing continuum. This effect, also known as the Pronik-Chuvaev effect after the authors who first demonstrated this effect using long-term monitoring of Mrk 6 \citep{pronik_chuvaev1972}, has been also observed and studied in a handful of nearby sources, e.g. NGC 4051 and NGC 4151 \citep[][and references therein]{Wang2005,Shapavalova2008,gaskell2021}. In Section 2, we outline the multi-component spectral fitting procedure ((Figure \ref{fig:spectra})) that is then used to create the continuum and \hb{} light curves over a broad time range. In Section 3, we describe the photoionization modelling setup involving a carefully, observation-oriented parameter space study that includes accounting for the continuum source and the BLR physical conditions using constant density single cloud models. In Section 4, we describe our results confirming the saturation of \hb{} with increasing continuum luminosity that can successfully reproduced with our photoionization modelling. In addition, we are able to provide constraints on the BLR local densities and the location of the \hb{} emitting BLR, the latter agreeing with the \hb{} time-lags reported from the long-term reverberation mapping monitoring for this source. In Section 5, we summarize the results and discuss some relevant issues pertaining to our study. Throughout this work, we assume a standard cosmological model with $\Omega_{\Lambda}$ = 0.7, $\Omega_{m}$ = 0.3, and H$_0$ = 70 \kms{} Mpc$^{-1}$.

\section{Spectral variability in NGC 5548}\label{sec2}

\subsection{Luminosity-luminosity correlations}

\begin{figure}
    \centering
    \includegraphics[width=\columnwidth]{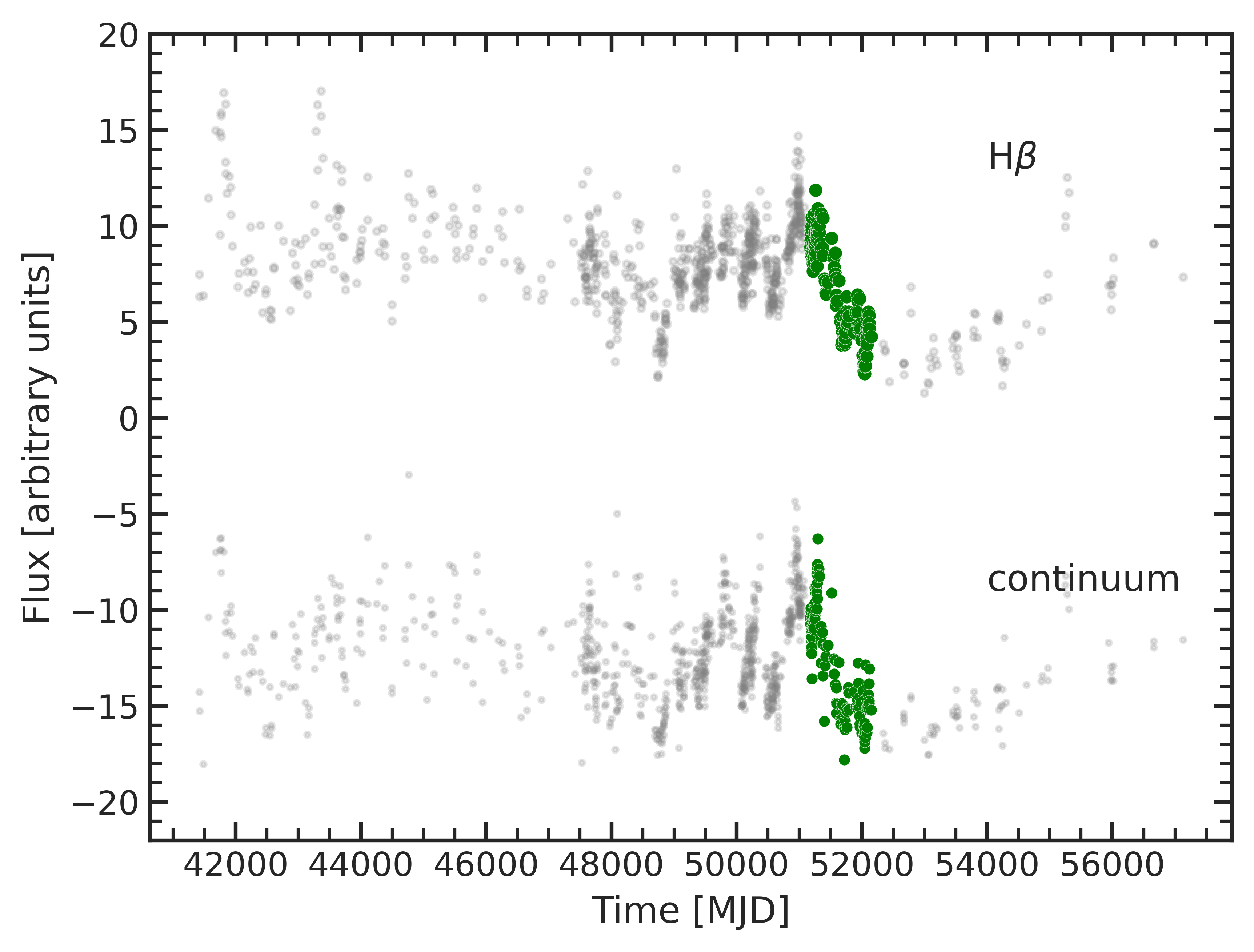}
    \caption{\hb{} flux and 5100\AA\ continuum (\lcont{}) flux light curves for NGC 5548 extracted from spectroscopic multi-component fitting. The range between 51170-52174 MJD used in this work is highlighted in dark green for both the \hb{} and \lcont{}.}
    \label{fig:lighct_curve}
\end{figure}

We extract the information of the continuum (at 5100\AA) and the \hb{} emission line\footnote{The \hb{} line luminosity is estimated from the integrated flux of the broad \hb{} line profile.} from the long term monitoring of NGC 5548.  In this paper, we focus on the observational light curves between the time range (in Modified Julian Days) 51170-52174 (see Figure \ref{fig:lighct_curve}). The choice of the time range is supported by the broad range covered by the \hb{} and the continuum luminosity that enables us study the extent of the flattening/shallowing trend observed in luminosity-luminosity correlation diagram shown in Figure \ref{fig:hbeta_v_inci_data}. The analysis incorporating the full observed range will be presented in a forthcoming work wherein we notice this shallowing effect in each epochs.

Spectroscopic observations were used to measure the \hb{} and continuum at 5100\AA~ fluxes. The spectra were compiled using AGN Watch \cite[see,][and the references within]{Petersonetal2002} and SAO monitoring program \citep{Shapovalovaetal2004} as described in \cite{bonetal2016,bonetal2018}. Spectral decomposition is obtained using ULySS software\footnote{available at \href{http://ulyss.univ-lyon1.fr}{http://ulyss.univ-lyon1.fr}}, full spectrum fitting technique \citep{Koleva2009,Bon14,Bonetal20}. The ULySS code is adopted for the purpose of Type-1 Seyfert spectra analysis to include necessary components of the AGN spectra - power law AGN continuum, nebular continuum, emission lines and \feii{} pseudo-continuum that are fitted simultaneously in order to minimize the effects of dependencies between parameters of the model \citep[for more details see,][]{Bon14,Bonetal20}. The host galaxy parameters were obtained using the low state spectra, where the contribution of the host absorption lines are more prominent, and these parameters were assumed to be the same for the rest of the monitoring spectra. The broad emission lines were modeled assuming several Gaussian components: two broad and one very broad. The details of the spectral decomposition is presented in \cite{Bon14,bonetal2016,bonetal2018,Bonetal20}.

The extracted light curves for the \hb{} and the 5100 \AA\ continuum luminosity from the spectral fitting are shown in Figure \ref{fig:lighct_curve}. Figure \ref{fig:hbeta_v_inci_data} shows the observed correlation between the continuum luminosity at 5100 \AA\ and the broad \hb{} emission line luminosity for the observed data within the given epoch, where each data point was observed spectroscopically. The continuum luminosity and the emission line luminosities for \hb{} were estimated from the corresponding fluxes obtained from the spectral decomposition and scaled based on the prescription used by the AGN STORM campaign \citep[see][]{Petersonetal2002,derosaetal2015}. We assume a luminosity distance of $\sim$ 75.01 Mpc (or $\approx$2.21$\times 10^{26}$ cm) obtained from the Two Micron All Sky Redshift Survey or 2MASS \citep{crooketal2007} to scale the fluxes to the corresponding luminosities. 


Looking carefully at Figure \ref{fig:hbeta_v_inci_data}, we can appreciate the gradual lowering of the gradient of the trend between the continuum and the \hb{} emission (``the taming of the derivative"). We have overall 153 spectroscopically estimated flux measurements concomitantly for the continuum and the emission lines from this observed time range. To assess the trend, we fit the observed trend with a 2$^{\rm{nd}}$ order polynomial regression fit of the form:

\begin{equation}
    L_{\rm{H\beta}} \;[\mathrm{erg\, s}^{-1}] \approx 1.097 \cdot 10^{41} + 0.034 L_{5100} - 3.885 \cdot 10^{-46} L_{5100}^2
\end{equation}

that is able to reproduce the underlying trend quite well.  The best-fit is supplemented with a 95\%\ confidence interval generated from bootstrapping 1,000 realizations of the observed data values. A first-order polynomial (linear) best-fit is also shown in Figure \ref{fig:hbeta_v_inci_data} for comparison. Although the linear-fit gives comparable values to the 2$^{\rm{nd}}$ order polynomial regression fit  and agrees with the low-luminosity median-binned intervals, it is unable to account for the shallowing in the correlation occurring at the higher luminosity values. This shallowing is well reproduced well with the 2$^{\rm{nd}}$ order polynomial regression fit. Previous works have also shown such shallowing of the correlation between the Balmer lines and the continuum luminosity in other known sources (Mrk 6 \citealt{pronik_chuvaev1972}, NGC4051 \citealt{Wang2005}, NGC4151 \citealt{Shapavalova2008}). We do have some scatter in the observed correlation and it's origin can be attributed to (a) significant delay in the response of the \hb{} to the continuum which has been noted in previous works \citep[e.g.][]{Cackett_Horne_2006,Goad_Korista_2014,Goad_Korista_2015,gaskell2021}, (b) imperfections of fitting of the spectra, and especially for low SNR spectra where the noise and resolution can underestimate the host contribution. This can lead to the AGN continuum appearing larger than it actually is.

\subsection{Spectral evolution of \hb{} and underlying AGN continuum in NGC 5548}

\begin{figure}
    \centering
    \includegraphics[width=\columnwidth]{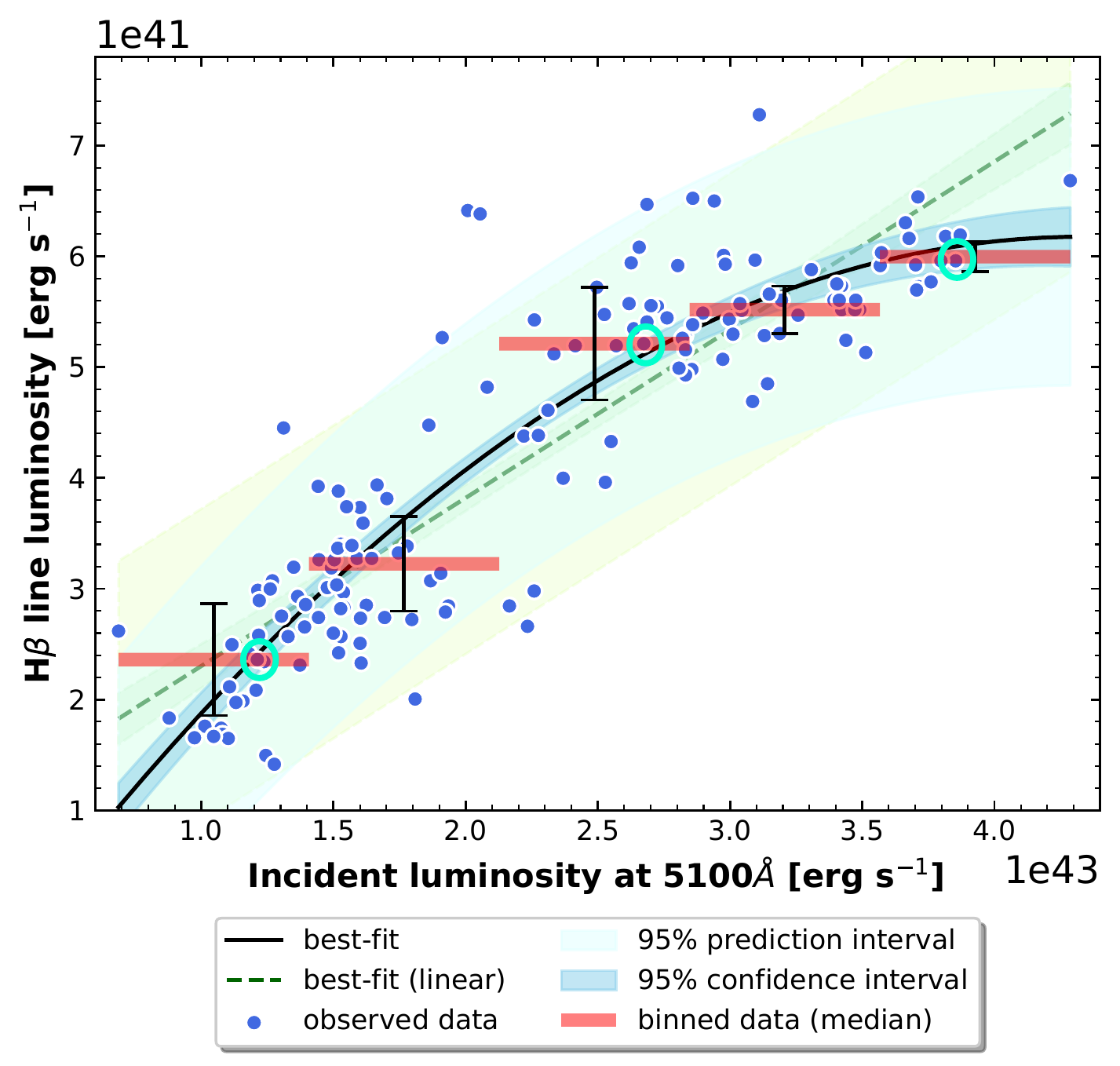}
    \caption{Luminosity-luminosity plot between continuum at 5100\AA\ vs \hb{} emission for the time range (in MJD) 51170 - 52174. The data points (blue dots) are fitted with a 2$^{\rm nd}$ order polynomial regression fit (black solid curve). 95\% confidence and prediction bands are shown (cyan and light-blue shaded regions, respectively) generated from bootstrapping 1000 realizations of the data. A first-order fit (green dashed line) is also shown for reference with respective confidence and prediction bands. The data is median-binned in 5 intervals (the extent of the bins are shown with horizontal red bars) to highlight the shallowing of the trend. The vertical errorbars (in black) mark the inter-quartile range for the corresponding bins. Spectra for the circled dots representing the three data-points from respective intervals are shown in Figure \ref{fig:compare_spectra}.}
    \label{fig:hbeta_v_inci_data}
\end{figure}

\begin{figure}
    \centering
    \includegraphics[width=\columnwidth]{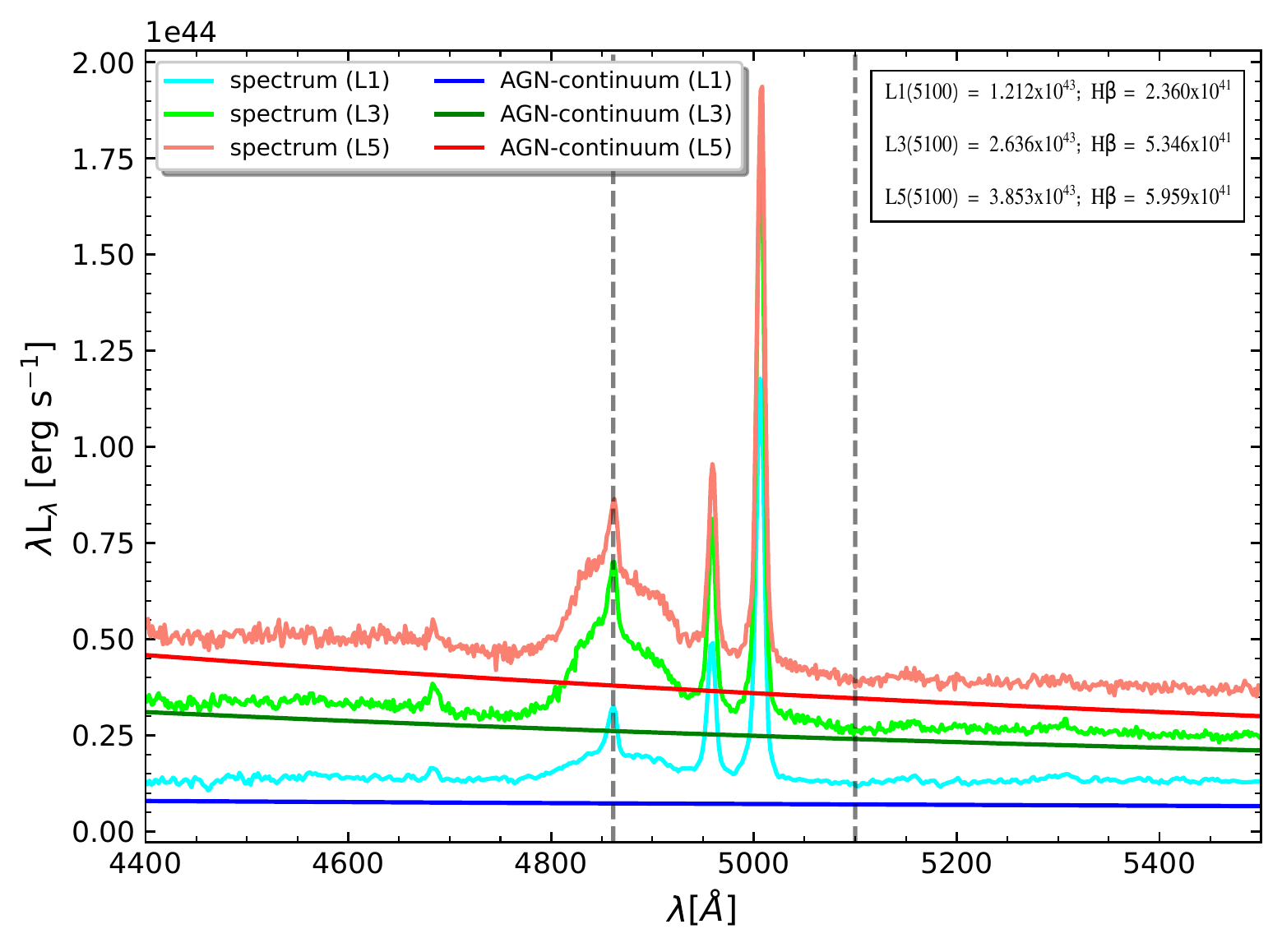}
    \caption{Comparison of the observed spectra and the corresponding AGN power-law continua for the three states as shown in Figure \ref{fig:spectra}. The spectra and the AGN continua are scaled to the actual luminosities that is in accordance to Figure \ref{fig:hbeta_v_inci_data}. The values for the \lcont{} and the corresponding \hb{} luminosities are shown in the inset on the upper right corner. The two vertical dashed lines mark the position of the rest-frame \hb{} emission line and the 5100\AA, respectively. The latter marks the location from where the continuum luminosity is estimated for each spectrum.}
    \label{fig:compare_spectra}
\end{figure}

Another way to appreciate the gradual saturation effect on the \hb{} as a function of increasing 5100\AA~continuum luminosity (\lcont{}) is shown in Figure \ref{fig:compare_spectra}. We extract the fitted model spectra and corresponding underlying AGN continua  and show in Figure \ref{fig:compare_spectra} for three instances, one from each luminosity state (Among these 5 states, we consider the lowest (L1, log \lcont{} $\sim$ 43.02), intermediate (L3, log \lcont{} $\sim$ 43.40) and the highest (L5, log \lcont{} $\sim$ 43.59) states) extracted from Figure \ref{fig:hbeta_v_inci_data}. The luminosity scale shown in this figure is in absolute units without any scaling between the three spectra, thus their changing behaviour can be appreciated directly. We provide the \lcont{} measured at 5100\AA~ and the corresponding \hb{} luminosity (at 4861.33\AA) for each of the three luminosity states in the inset of Figure \ref{fig:compare_spectra}. Comparing the slopes\footnote{the slopes are estimated by measuring the ratio of the difference between the \hb{} luminosities to the difference between the corresponding \lcont{}. In other words, the slope, m = $\frac{\rm L^{'}_{H\beta} - L_{H\beta}}{\rm L^{'}_{cont} - L_{cont}}$, where L and L' represent the two luminosity states.} of the \hb{} vs. \lcont{} correlation between the consecutive luminosity states, we have: (a) m$_{\rm L1\rightarrow L3}$ = 0.021, and (b) m$_{\rm L3\rightarrow L5}$ = 0.005, confirming the shallowing of the trend between the \hb{} luminosity with respect to the increasing 5100\AA~ continuum luminosity.



\section{Photoionization computations}\label{sec3}

\begin{figure}[hbt!]
    \centering
    \includegraphics[width=\columnwidth]{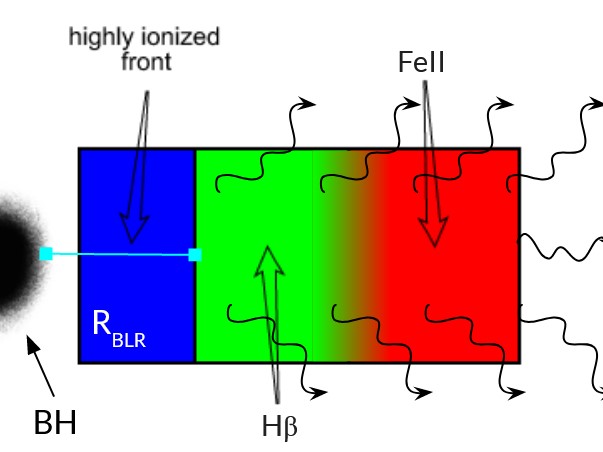}
    \caption{Illustration of a single cloud model setup in \textmyfont{CLOUDY} used in this work. The black hole (with the continuum emitting accretion disk included, not shown here) is shown at the left of the illustration. The broad-line region (BLR) is depicted in a plane parallel configuration with fully and partially-ionized zones. The distance of the BLR from the central continuum source is marked at the onset of the \hb{} emitting zone as characterized by the reverberation mapping estimates. The sketch is not drawn to scale: the geometrical depth of the emitting cloud is greatly expanded to show its internal structure.}
    \label{fig:cloud}
\end{figure}

In order to assess and reproduce the observed trends for the \hb{} emission line with respect to the continuum luminosity, we perform a suite of \textmyfont{CLOUDY} \citep{f17} photoionization simulations. We consider a grid of single cloud models covering a wide range in the parameter space that includes (a) the continuum luminosity, (b) the distance between the continuum source and the onset of the BLR cloud (\rblr{}), (c) shape of the ionizing continuum characterized by the spectral energy distribution (SED) appropriate for NGC 5548, (d) the BLR local density, and (e) the geometrical depth of the BLR clouds parametrized by the cloud column density. We refer the readers to an illustration of a single BLR cloud in Figure \ref{fig:cloud} that is photoionized by the continuum radiation emanated in the inner sub-parsecs from the BH. The BLR cloud is stratified and the subsequent separation into zones of fully ionized and partially ionized regions is shown in this figure for reference highlighting the location of the emission of the \hb{} and other low-ionization emission lines, such as \feii{}, confirmed by prior reverberation mapping estimates and photoionization models \citep[see][for more details]{panda18b}. The first three parameters are constrained from prior observations, e.g., we consider five luminosity states that are obtained by median-binning the full range of the observed data (see the horizontal red bars in Figure \ref{fig:hbeta_v_inci_data}). These luminosity states are then provided as one of the inputs to the set of simulations. It is worth noting here that the black vertical bars for each bin represent the inter-quartile (between the 25$^{th}$ and 75$^{th}$ percentiles) range for the corresponding bin that show the extent of the dispersion in each bin. The radial distance between the source and the BLR cloud (\rblr{}) is set by the distance estimated from the time lags between the continuum and emission line light curves obtained from prior reverberation mapping studies carried out for NGC 5548 \citep[][and references therein]{horneetal2021}. \citet{horneetal2021} estimated the range for this \hb{}-based \rblr{} to be as short as 5 light days\footnote{1 light day = 2.59$\times 10^{15}$ cm.} and extending up to 40 light days. We thus consider a \rblr{} range between 10$^{16}$-10$^{17}$ cm keeping consistency with these studies. Next, we require to know explicitly the shape of the SED that then allows to scale the monochromatic continuum luminosity in order to estimate the total number of ionizing photons that effectively lead to the line production in the gas-medium, in our case, the BLR. We assume the SED provided in \citet{dehghanianetal2019} used in AGN STORM modeling of NGC 5548. This SED was originally shown in \citet{mehdipouretal2015} to which later a low- and a high-energy data point were added to extend the coverage of the SED on either side by \citet{dehghanianetal2019}. This SED is available directly from the \textmyfont{CLOUDY} database\footnote{\href{https://gitlab.nublado.org/cloudy/cloudy/-/blob/master/data/SED/NGC5548.sed}{https://gitlab.nublado.org/cloudy/cloudy/-/blob/master/data/SED/NGC5548.sed}}. 
The range for the remaining two physical parameters - the BLR local density and BLR cloud sizes, are set based on previous studies \citep{panda_etal2020, korista_goad_2000}. The local BLR density is considered within the range $10^{9} \leq n_{\rm{H}} \leq 10^{13}\;(\rm{cm^{-3}}$). This range for the local BLR density is considered following the conclusions from prior studies that also focused on reproducing the low-ionization emission including \hb{} - using emission line ratios that trace the local density and other BLR diagnostics for a sample of quasars \citep{negrete_etal2013, sniegowska_etal2021}, and from photoionization modelling estimates \citep{panda_etal2020, panda2021}. These studies have shown that the considered range reproduce well the range of the ionization parameter (\textit{U}), given that the BLR distance from the continuum source is set from the observational \RL{} scaling relation \citep{bentz_etal2013, horneetal2021}. Density values any higher or lower curtail the emission from these low-ionization lines. If the  ionization parameter is too low, the ionizing continuum does  not provide sufficient energy in (i.e., few photons carrying energies $\gtrsim$ 1 Rydberg). If too high,  the parent ionic species become higher than the ones from the observed emission lines. 

The BLR cloud sizes are determined by assuming that the line-emitting region of the BLR are sufficiently optically thin, i.e. with optical depths ($\tau$) $\sim$ 1-2, such that any effect from electron back-scattering from the surface or inside of the cloud leading to additional cooling of the region is low and can be safely neglected. The cloud column density ($N_{\rm{H}}$) command in \textmyfont{CLOUDY} allows the user to set the size ($d$) of the cloud given by the relation: $d = \frac{N_{\rm{H}}}{n_{\rm{H}}}$, where $N_{\rm{H}}$ and $n_{\rm{H}}$ have their usual meaning. Here, we consider two cases for the cloud column density, $N_{\rm{H}} (\rm{cm^{-2}}$) = $10^{22}$ and $10^{23}$. These two values cover the range of neutral gas column density derived from X-ray observations, $10^{22} - 10^{23}\; \rm{cm^{-2}}$\ \citep[e.g.,][]{waltercorvoisier89,netzer90,singhetal91}. For NGC 5548, a multi-epoch study based on simultaneous coverage in the soft and hard X-ray domain consistently yields $ N_{\rm{H}} \approx 9 \cdot 10^{22}$  $\rm{cm^{-2}}$ \ \citep{ursinietal15}.

This 1 dex range in $N_{\rm{H}}$ is sufficient to test the effect of the increasing BLR cloud sizes in optimizing the \hb{} emission from the BLR and is consistent with prior studies involving the analysis of NGC 5548 \citep{dumont_etal1998,korista_goad_2000}. In this work, we consider solar abundances (Z = \zsun{}) for the composition of the BLR cloud. This assumption of the composition is corroborated by previous studies for NGC 5548 \citep{dehghanianetal2019} and is thought to be usually appropriate  for sources like NGC 5548 that belong to Population B \citep{punslyetal18a}. The Population B sources are characterized by relatively `broader' broad emission lines, show higher variability amplitude in their light curves and produce relatively low \feii{} emission compared to their counterparts, i.e. the Population A sources \citep{mar18,panda19b}. The modelled luminosities for the \hb{} are extracted from these photoionization models.

\section{Trends and Analyses}\label{sec4}

\begin{figure*}
    \centering
    \includegraphics[width=2\columnwidth]{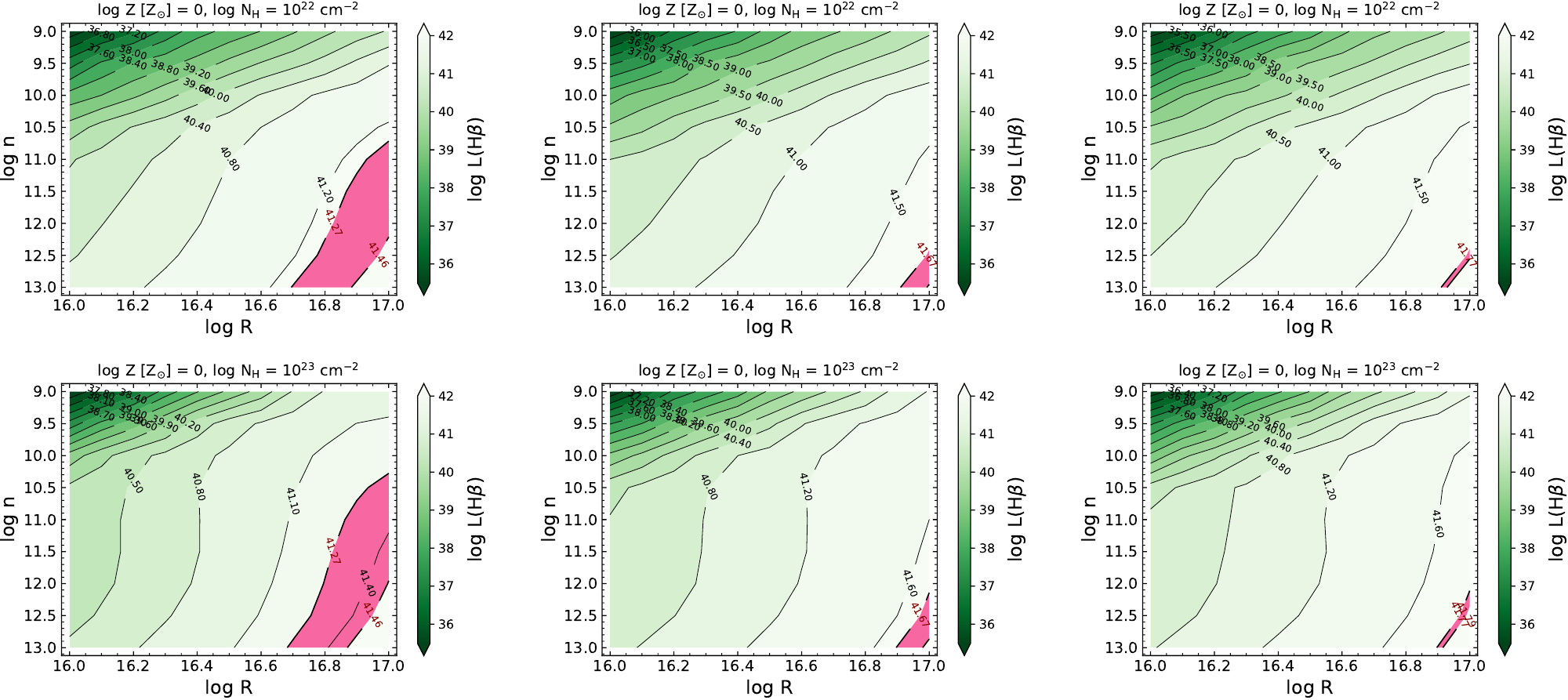}
    \caption{log R - log n histogram matrix generated using the \textmyfont{CLOUDY} simulations. The auxiliary axis is shown as a colormap representing the luminosity of the \hb{} emission line (in log-scale) which is also represented using black contours in each panel. A covering factor of 20\% is used to re-scale the luminosities. The panels (from LEFT to RIGHT) represent increasing continuum/incident luminosity (in ergs s$^{-1}$, in log-scale): (LEFT) 43.02, (MIDDLE) 43.40, and (RIGHT) 43.59. The panels (TOP and BOTTOM) represent increasing broad-line region cloud column density (in cm$^{-2}$, in log-scale): (TOP) 22 and (BOTTOM) 23. Each of the simulations were modelled using a composition with solar abundances. The input spectral energy distribution for the simulations is taken from \citet{dehghanianetal2019}. The red shaded region in each panel marks the inter-quartile range for the corresponding median-binned data as per Figure \ref{fig:hbeta_v_inci_data}.}
    \label{fig:cloudy_hb}
\end{figure*}

The full, modelled parameter space considered in our photoionization computations are shown in Figure \ref{fig:cloudy_hb}, where the X- and Y- axes represent the extent of the BLR distance from the continuum source and the local BLR density, respectively. The auxiliary axis shown as a colormap represents in this case the luminosity of the \hb{} emission line (in log-scale). Figure \ref{fig:cloudy_hb} is shown as a 3$\times$3 matrix of 2D density histograms with varying continuum luminosity states (from left to right) extracted from the \textit{flattening} curves that are median-binned into 5 distinct luminosity states as described in Section \ref{sec2}. Among these 5 states, we consider the lowest (log \lcont{} = 43.02), intermediate (log \lcont{} = 43.40) and the highest (log \lcont{} = 43.59) states to infer the gradual changes associated with the corresponding line emission (see also Figure \ref{fig:compare_spectra} for reference). The effect of the increasing column density, going from the relatively low ($N_{\rm{H}}$ = 10$^{22}$) to high ($N_{\rm{H}}$ = 10$^{23}$),  are shown going from top to bottom in the figure. \textmyfont{CLOUDY}, by default, provides the luminosities assuming a 100\% covering factor (CF), i.e. solid angle that is equal to 4$\pi$ steradians. We utilize a covering factor of 20\% that is found to be viable for a near-Keplerian distribution of the BLR clouds \citep{korista_goad_2000, baldwin_etal2004, sarkar_etal2021, panda2021}. 


In order to assess the performance of the photoionization computations against the observed values obtained from the \textit{flattening} curve, we supplement the 2D histograms with the corresponding estimation of the \hb{} line luminosities that are shown using the red shaded regions in each panel of Figure \ref{fig:cloudy_hb}. The extent of the shaded region is set by the corresponding inter-quartile range as shown on the \textit{flattening} curve, i.e. Figure \ref{fig:hbeta_v_inci_data}.

An expected result that is observed from the upper panels ($N_{\rm{H}}$ = 10$^{22}$) in Figure \ref{fig:cloudy_hb} is that the local density is confined to values that are around $10^{11}$ cm$^{-3}$ or higher. The characteristic BLR density has been explored and substantiated in many previous studies which include the modelling of the emission from the low-ionization lines, including \hb{} \citep[e.g.][]{korista_goad_2000,martinez-aldama_2015,panda18b,panda19b,panda_etal2020,panda2021}. These ($10^{11}$ cm$^{-3}$) are the typical densities obtained from the diagnostic ratios in the UV regime of spectra - broad inter-combination and permitted lines like C III$\lambda$1909/Si III]$\lambda$l1892 and Al III$\lambda$1860/Si III]$\lambda$1892 which correspond to the densest emitting regions that are likely associated with low-ionization lines’ production \citep[e.g.][]{baldwin_etal2004,matsuoka08}. The range of the local density shifts towards higher values as the continuum luminosity is increased (going from left to right). We notice a similar behaviour in the bottom panels ($N_{\rm{H}}$ = 10$^{23}$ \ cm$^{-2}$). 

The tapering of the width of the shaded region as we increase the continuum luminosity is due to the lowering of the dispersion in the corresponding luminosity bin that is highlighted by the smaller inter-quartile range. We can notice the gradual shift in the shaded region towards higher values of \rblr{} when we increase the continuum luminosity. This result is in-line with the prediction from the \RL{} relation \citep{bentz_etal2013}. On the other hand, we notice that the \rblr{} that correspond to the shaded region is rather limited and doesn't vary much with the increase in the column density.

We also test a case with higher covering fraction, i.e., at 50\% for all the cases mentioned above (see Figure \ref{fig:cloudy_hb2}) that is still within the realms of the solutions as per \citet{korista_goad_2000}. The only difference in their approach was that the models were computed using the locally optimized cloud (LOC) model \citep{baldwin1995} while here we limited our approach to constant density single cloud models to check for the dependencies on the various physical parameters as a first test. A detailed analysis incorporating the LOC model will be presented in a forthcoming paper. The solutions obtained for the 50\% covering fraction cases have a rather loose constraint on the local density - the density range gradually shifts towards lower values with increasing column density. 


\section{Conclusions}
In this paper, we test the connection between the continuum variability in the optical regime (5100\AA) and the corresponding \hb{} response to it for NGC 5548. We perform a multi-component spectral fitting to optical spectra observed in the time range between 51170-52174 (in MJD) and realise the shallowing of the trend between the \hb{} and the corresponding 5100\AA~ AGN continuum luminosities. We perform a suite of photoionization models assuming constant density, single cloud approximation using \textmyfont{CLOUDY} in order to infer this observed trend. The flattening of the luminosity-luminosity relation between the continuum and the \hb\ emission line in NGC 5548 is consistent with the presence of a  medium of moderate local density, moderate cloud column density whose response diminishes with increasing continuum luminosity. Below, we summarize our findings in this paper:

\begin{enumerate}
    \item The saturation of \hb{} luminosity as a function of the rising AGN continuum luminosity in NGC 5548 is evident, and follows (within scatter) a 2$^{\rm nd}$ order polynomial.
    \item Taking into account the parameters consistent for photoionization modelling the BLR in NGC 5548, we obtain a marked agreement between the observed emission and the photoionization models.
    \item We are able to constrain the local density of the BLR as a function of the rising continuum luminosity for a BLR covering factor (CF) ~ 20\%. For higher CFs, the densities are rather unconstrained.
    \item We confirm a systematic shift (outwards) in the estimated BLR size (\hb{}-based) with increasing AGN continuum luminosity in our models.
    \item In addition, the cloud column sizes required to observe the saturation of the \hb{} emission should subsequently have cloud column densities within 10$^{22}$-10$^{23}$ cm$^{-2}$.
\end{enumerate}

The behavior observed in NGC 5548 might be typical for Population B sources  radiating at relatively low \ledd\ ($\lesssim$ 0.2), where a large amount of gas might not be available, small \ledd{}. Sources radiating at high Eddington ratio are apparently much more stable photometrically \citep{duetal18}. Indeed, there are other Population B sources for which the  shallowing trend in the \hb{} broad-line vs. continuum luminosity has been reported (Mrk 6 \citealt{pronik_chuvaev1972}, NGC4051 \citealt{Wang2005}, NGC4151 \citealt{Shapavalova2008}). We intend to study and constrain the physical conditions in the BLR for these sources in addition to covering the full observed range for NGC 5548, inclusive of the ``BLR holiday'' time interval \citep{goad_etal_2016,pei_etal_2017,dehghanianetal2019}, with the modelling approach demonstrated here in an upcoming work.

\section*{Acknowledgments}

We would like to thank Prof. Martin Gaskell and Prof. Robert Antonucci for fruitful discussions and suggestions. SP acknowledges the financial support from the National Science Centre, Poland grant No. 2017/26/A/ST9/00756 (Maestro 9). NB and 
EB acknowledge the support of Serbian Ministry of Education, Science and Technological Development, through the contract number 451-03-68/2020-14/200002.

\subsection*{Author contributions}

S. Panda performed photoionization computations and statistical analyses, and put together the text and the figures. E. Bon provided the idea for the project and with N. Bon performed the spectral analysis of the source's spectra and compiled the catalogue with relevant information and assisted with the text. P. Marziani assisted with the overall analysis and text. 

\subsection*{Financial disclosure}

\fundingAgency{National Science Centre, Poland} grant No. \fundingNumber{2017/26/A/ST9/00756 (Maestro 9)}, \fundingAgency{Brazilian National Council for Scientific and Technological Development} \fundingNumber{(CNPq) Fellowship 164753/2020-6}, and \fundingAgency{Serbian Ministry of Education, Science and Technological Development} contract number \fundingNumber{451-03-68/2020-14/200002}.

\subsection*{Conflict of interest}

The authors declare no potential conflict of interests.





\appendix

\section{Modelled \hb{} luminosity maps for a higher covering fraction\label{app1}}

\begin{figure*}
    \centering
    \includegraphics[width=2\columnwidth]{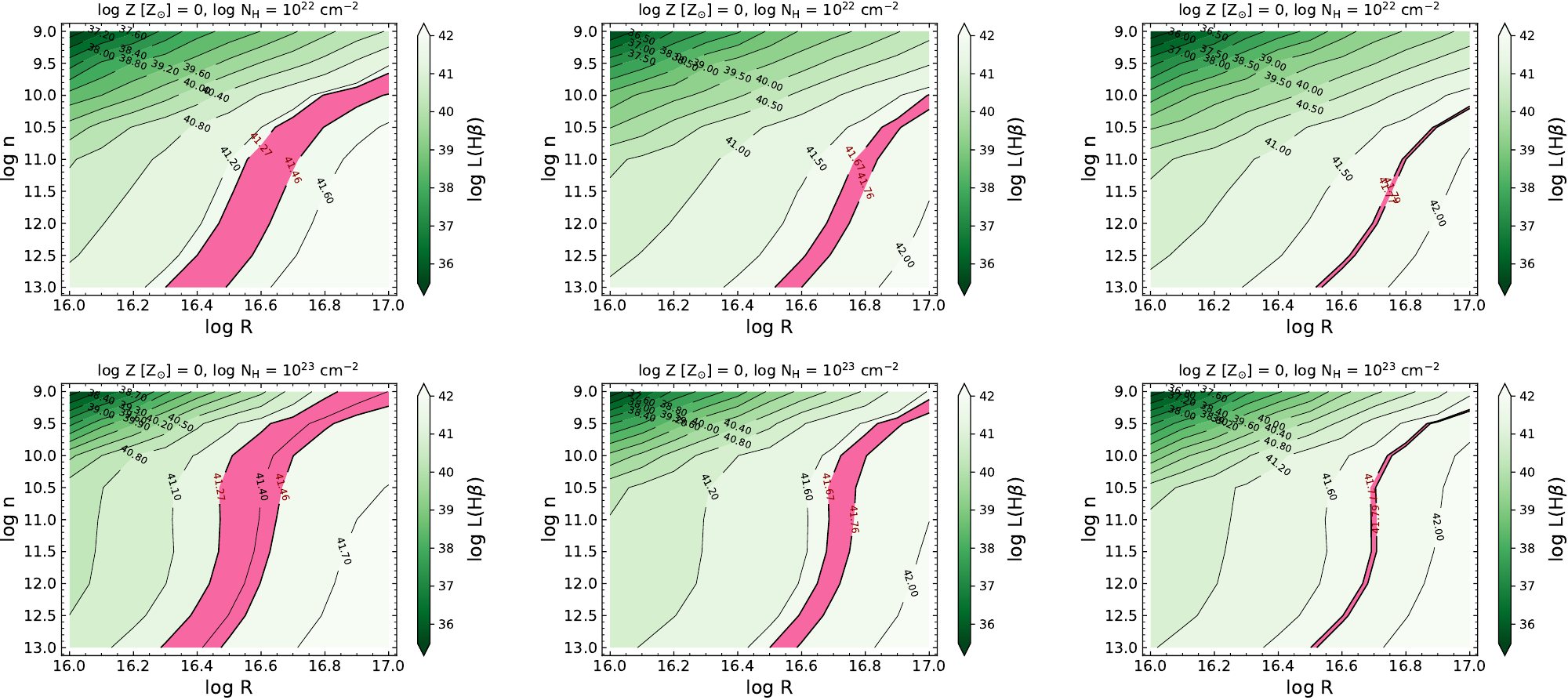}
    \caption{Similar to Figure \ref{fig:cloudy_hb} but a covering factor of 50\% is used to re-scale the luminosities.}
    \label{fig:cloudy_hb2}
\end{figure*}




\nocite{*}
\bibliography{references}%

\section*{Author Biography}

\begin{biography}{\includegraphics[width=70pt,height=70pt]{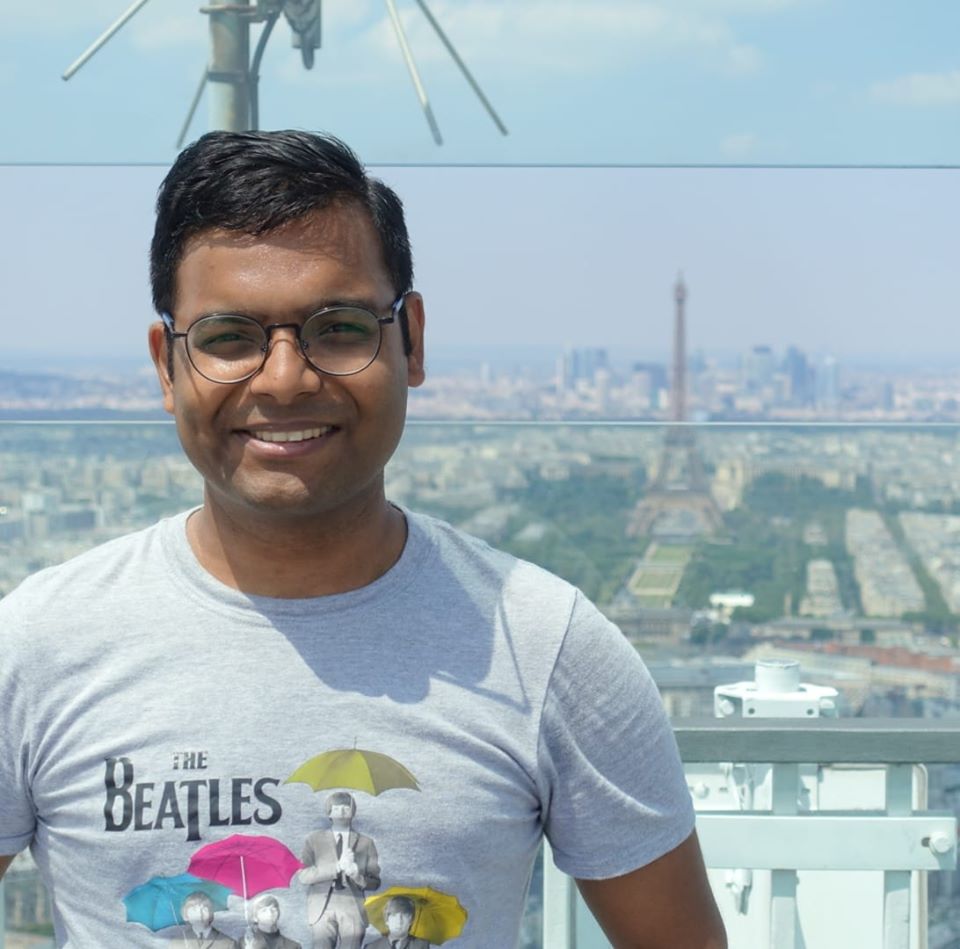}}{\textbf{Swayamtrupta Panda.} Swayamtrupta Panda, PhD., finished bachelors studies in Physics at the National Institute of Technology in Rourkela, India in 2014, after defending the bachelor thesis ``Physics of Seyfert Galaxies'' supervised by Prof. Retd. Prajval Shastri (Indian Institute of Astrophysics, Bengaluru, India) as a INSA-NASI-IASc. research fellow. Subsequently, he continued with the master degree in Physics with specialization in Astronomy and Astrophysics with the thesis ``GALEX Observations of Planetary Nebulae'', supervised by Prof. Ananta Charan Pradhan, which he defended at the National Institute of Technology in Rourkela, India in 2016. In the meantime, he held a MITACS research fellowship position (2015) at the University of Victoria (British Columbia, Canada) under the supervision of Prof. Sara E. Ellison where he studied impact of galaxy mergers on AGN triggering using all-sky surveys and cosmological simulations. In the fall of 2016, he started doctoral studies jointly at the Center for Theoretical Physics and the Nicolaus Copernicus Astronomical Center of the Polish Academy of Sciences, under the supervision of Prof. Bo\.zena Czerny. In September 2021, he defended the PhD thesis ``Physical Conditions in the Broad Line Regions in Active Galaxies''. In the fall of 2021, he will begin his postdoctoral position at the Kavli Institute for Astronomy \& Astrophysics in Prof. Luis Ho's group as a Kavli-Boya Fellow.}
\end{biography}

\end{document}